\documentstyle[aps,pra]{revtex}
\draft
\begin{document}
\twocolumn[\hsize\textwidth\columnwidth\hsize
\csname @twocolumnfalse\endcsname

\title{\begin{flushright}
{\small EHU-FT/9901, quant-ph/9905023}
\end{flushright}Free motion time-of-arrival operator and probability
distribution}
\author{I.L. Egusquiza$^1$ and J.G. Muga$^2$}
\address{$^1$Fisika Teorikoaren Saila,
Euskal Herriko Unibertsitatea,
644 P.K., 48080 Bilbao, Spain\\
$^2$Departamento de F\'\i sica Fundamental II, Universidad de
La Laguna, La Laguna, Tenerife, Spain}
\date{\today}

\maketitle

\begin{abstract}
We reappraise and clarify the contradictory statements found in the
literature concerning the time-of-arrival operator introduced
by Aharonov and Bohm in Phys. Rev. {\bf 122}, 1649 (1961).
We use Naimark's dilation theorem to reproduce the
generalized decomposition of unity (or POVM) from any self-adjoint
extension of the operator, emphasizing a natural one,
which arises from the
analogy with the momentum operator on the half-line. General time
operators are set within a unifying perspective.
It is shown that they
are not in general related to the time of arrival, even though they may have 
the same form.

\end{abstract}

]

\section{INTRODUCTION}
Even though there is no question that experimentalists measure
distributions of time of arrival for quantum systems, there has been
a long debate about the capabilities of standard quantum mechanics to
address the very concept of time
as an observable.
In a nutshell, the problem arises because we expect observables to be
represented in the quantum formalism by self-adjoint operators,
whereas an old theorem of Pauli \cite{pauli} states that for a
semibounded self-adjoint operator $\hat H$ no conjugate self-adjoint
operator $\hat T$ can exist, i.e., no operator that is self-adjoint
and satisfies the canonical commutation relation $[\hat H,\hat
T]=i\hbar$ over a dense domain. In other words, there is no
self-adjoint time operator if the Hamiltonian is bounded from below,
as we normally expect it to be.
However the theorem has not discouraged theorists from
attempting to fit such an immediate classical concept into the
standard framework (see \cite{microst} for a general discussion). In
relatively recent times a rather
satisfactory answer has
been given for the arrival time of freely moving states in one dimension,
from the point of view of generalized spectral functions,
or positive-operator-valued measures (POVM)
\cite{ludwig,holevo,bglbook,bglpaper}. In short, the relevant
statement is that self-adjointness is not strictly necessary for the
formulation of probabilities and the reality of measurements of
observables, and that, in fact, any given observable is uniquely
characterized (in a specific sense) by the probability distributions
of measurement results in the different states accessible to the
system (for a much more developed presentation, refer to
\cite{holevo,bglbook}).

This result has sparked a renewed interest in the analysis of old and
new proposals for time operators in general, and in particular for the
time-of-arrival operator
$\hat T_{AB}$ (see below) introduced by Aharonov and Bohm \cite{ahb}, as in
\cite{bglpaper,allcock,paul,razavi,kijold,kijnew,werner,%
gianni,delmu,grt,oru,toller,juan,mlppra,pegg}.

However, a number of basic issues are still in dispute, or in need of
clarification. In particular, contradictory statements
may be found about the domain of the time operator $\hat T_{AB}$
\cite{paul,gianni},
or about its eigenfunctions \cite{grt,gianni}; and the
deficiency indices, which are important to
determine its self-adjoint extensions, appear to be,
following different authors,
either (0,1) or (0,2) \cite{razavi,gianni},
but only one of the possibilities may be correct.
Even though several self-adjoint ``variants'' of $\hat T_{AB}$ have been proposed,
\cite{kijold,grt,delmu} none of them is a self-adjoint extension
of $\hat T_{AB}$ in a proper sense. The framework to consider
these extensions is
Naimark's dilation theorem \cite{holevo,bglbook,akgl}, but an
explicit simple construction of a self-adjoint
extension is lacking (For a rather abstract presentation of the
theorem applied to the time-of-arrival problem, under the guise of
``screen observables'', see \cite{werner}; also
\cite{holevo,bglpaper,bglbook}).
In addition, the form of the time operators
in different representations
may also be, and has been, misleading: the same operator has been
presented in different forms, but different operators have also been
written in the same form.

It appears that in order to make
further progress in this field without
being trapped by any of the numerous pitfalls mentioned,
a detailed analysis of these topics is required, and
their clarification is the main objective of this paper.
Even though some of the arguments have to be
rather technical, our aim is to provide an account readable
by a general physicist.
We shall in particular
introduce a description of the extension theorems as they
apply to the problem at hand, and
develop a general analogy, sketched by Holevo \cite{holevo},
between the momentum operator in the half
line and the time operator. We hope
that it
will provide a much better understanding of the relevant issues.
It will also allow
to join the strands given by i) the initial
proposal of the time-of-arrival operator of Aharonov and Bohm
\cite{ahb},
its modifications or variants
\cite{kijold,delmu,grt,oru};
ii) the POVM idea as applied to
quantum physics and to the time operator in particular
\cite{bglbook,gianni}; iii) the axiomatic approach of Kijowski to
obtain good probability distributions for the time of arrival
\cite{kijold,kijnew}; and iv) some apparently formal constructions
in Allcock's seminal paper\cite{allcock}.

As is well known, the momentum operator in the half-line is
is not self-adjoint.  Moreover, it is maximally symmetric,
because its deficiency indices are $(1,0)$, and it admits no
self-adjoint extension on the Hilbert space of functions defined on
the half-line (for a fast and good review of deficiency indices and
the Neumann theory of extensions, with particular emphasis on the
physical examples of momentum operator and
Laplacian, see
\cite{galindo} - we will anyhow give a self-contained description in
section II).  On the other hand, there is a natural extension, namely,
the momentum operator on the full line.  It should be noticed that
this is most definitely not the unique extension, but just one of the
infinitely many possible.  Nonetheless, its
``naturalness'' is
well justified
from a physical point of view.  The question might then be posed as to
how are we to recover any information about the distribution of
probability of momentum in the half-line from the full line extension.
This is actually the content of Naimark's dilation theorem, together
with the theorem that asserts the uniqueness of the POVM associated
with a maximally symmetric operator: that the {\sl unique}
distribution of probability associated with the momentum operator on
the half-line is recovered from the spectral decomposition of the
self-adjoint operator momentum on the full line, by simple projection.
Equivalently stated, we can reproduce the {\sl physical} content of
the momentum operator on the half-line from the standard, well-known
techniques that apply to self-adjoint operators.

We will show that the different alternatives that have been proposed
to make sense of the time operator of Aharonov and Bohm are actually
consequences of this general result, that we have phrased in terms
of momentum operators, but which can be extended to other
cases as well.
The generality of the result also explains the coincidence of
different constructions, in particular the axiomatic approach of
Kijowski \cite{kijold,kijnew} and the attempts to create ex novo a
self-adjoint operator (\cite{delmu} is a relevant example).

We organize the paper as follows: in the next section we give a
complete and very basic description of the momentum operator on the
half-line and the problems that appear there. We also give the natural
extension and describe more fully Naimark's dilation theorem in this
context. In section III we establish the parallel with the
time-of-arrival operator of Aharonov and Bohm, to be followed in the
next section  with the
general case of a time operator. Section V is devoted to
discussing and restating
the different results known in the literature in the terms
used in the previous sections.

\section{THE MOMENTUM OPERATOR}

Consider the momentum operator $\hat p=-i\hbar\partial_{x}$ defined on
a dense domain of the Hilbert space of square integrable functions on
the half-line, ${\cal H}_{>}=L^{2}\left({\bf R}^{+},{\rm d}x\right)$.
More precisely, it is defined on $D(\hat p)$, the subspace of square
integrable absolutely continuous functions $\psi$ whose derivative is
also square integrable, and such that $\psi(0)=0$.  Clearly, this
operator is symmetric on its domain.  Let us write down a set of
generalized (weak) eigenfunctions that provide us with a resolution of
the identity.  The adequate set, which does not belong to the Hilbert
space and is parametrized by real $p$, is given by
$\psi_{p}(x)=\exp(ipx/\hbar)/\sqrt{2\pi\hbar}$, as is only to be
expected.  They form a complete basis, since
$\int_{-\infty}^{\infty}{\rm
d}p\,\overline{\psi_{p}(x')}\psi_{p}(x)=\delta(x'-x)$.  However,
they are not orthogonal (in the generalized sense):
\[\int_{0}^\infty{\rm d}x\,\overline{\psi_{p'}(x)}\psi_{p}(x)=
\frac12 \delta(p-p') +\frac{i}{2\pi}{\rm P}\frac1{p-p'}\,,\] where P
stands for principal part.  This is evidently different from the usual
constructions for self-adjoint operators.

Let us now pass on to the position operator $\hat x$, defined on
those square integrable functions on the half-line, $\psi\in{\cal H}_{>}$,
such that $\int_{0}^{\infty}{\rm d}x\,x^2|\psi(x)|^{2}<+\infty$. This
operator is self-adjoint on its domain. Additionally, it is true
that $[\hat x,\hat p]=i\hbar$ on a dense domain. However, in the case
at hand the position operator is bounded from below, and Pauli's
theorem therefore applies. It follows that $\hat p$ is not a
self-adjoint operator and admits no self-adjoint extension in ${\cal
H}_{>}$.

We shall now rephrase the statements of the previous two paragraphs in
a different way. The domain of the operator adjoint to $\hat p$,
i.e., $\hat p^{\dagger}$, according to von Neumann's formula,
is $D(\hat p)\oplus N(i)\oplus N(-i)$, where $N(\pm i)$ are
the spaces of eigenvectors of $\hat p^{\dagger}$ with eigenvalues
$\pm i$ respectively.
This comes about because, even though $\hat p$, being symmetric, has
no imaginary eigenvalues, its adjoint does have eigenvectors with
imaginary eigenvalues. As a matter of fact, the functions
$\exp(i\lambda x/\hbar)$, with $\lambda$ a complex number with
positive imaginary part, are all of them eigenvectors of $\hat
p^{\dagger}$ of eigenvalue $\lambda$. Notice that the dimension of
$N(\lambda)$, the space of eigenvectors with eigenvalue $\lambda$, is
the same for all $\lambda$ with positive imaginary part. This is
called the first deficiency index of the operator $\hat p$, while the
second one is the dimension of the space of eigenvectors for any given
eigenvalue of negative imaginary part, in our case, 0. That is, the
deficiency indices of $\hat p$ are $(1,0)$.
But self-adjoint operators
have deficiency indices $(0,0)$, as do essentially self-adjoint
operators, which, although not self-adjoint, have a unique
self-adjoint extension, namely, their closure. If we were to build a
self-adjoint extension
of
$\hat p$ we would need to include
somehow those elements of $D(\hat p^{\dagger})$ that are not in
$D(\hat p)$, but in such a way that the imaginary parts compensate.
When the deficiency indices are equal, this is achieved by the use of
a unitary transformation from $N(\bar\lambda)$ to $N(\lambda)$ (this
is the von Neumann theory of self-adjoint extensions, whereby each
self-adjoint extension in the Hilbert space of definition of an
operator over a dense domain, with equal defect indices, is given by
a unitary transformation between the deficiency subspaces). As
the deficiency indices in our case are different, it is not possible
to extend $\hat p$ to include an action over the whole of the domain
of $D(\hat p^\dagger)$ such that the imaginary parts compensate.
Since $\hat p$ is a closed operator,
symmetric over its domain, which is dense
in ${\cal H}_{>}$, and its deficiency indices are unequal, one of them
being 0, we say that
it is a maximally symmetric operator, and the previous result is that
it admits no self-adjoint extension over ${\cal H}_{>}$.

Retaking the complete set of generalized eigenfunctions $\psi_{p}(x)$,
we can construct a positive-operator-valued measure (POVM) $F$.
This is a map
from intervals in the real line (a $\sigma$-algebra
of subsets of a nonempty set)  to the positive
operators over a Hilbert space that satisfies three axioms that we
will illustrate with the example at hand. The map for our case is
given by
\[
\langle \phi|F([a,b])\psi\rangle =
\int_{a}^b{\rm d}p\int_{0}^{\infty}{\rm
d}x \int_{0}^{\infty}{\rm
d}y\,\overline{\phi(x)}\psi_{p}(x)\overline{\psi_{p}(y)}
\psi(y)\,,\]
over all $\phi,\psi\in{\cal H}_{>}$.  This map sends intervals of the
real line to positive operators acting on ${\cal H}_{>}$ (first
axiom), which add together when the intervals are disjoint (second
axiom), and which add to the identity operator when summed over the
real line, because of the completeness proved above (third axiom).
This differs from the usual decompositions for self-adjoint operators
in that it does not fulfill the property that the positive operators
be projectors, i.e., in our case $F([a,b])^2\neq F([a,b])$, because of
the lack of orthogonality shown above.

Now we have the POVM associated with $\hat p$, we can reconstruct the
operator and the probability distribution of its values. The action
of the operator is given by
\[(\hat p\varphi)(x)=\int_{-\infty}^{\infty}{\rm d}p\int_{0}^\infty{\rm
d}y\, p\,\psi_{p}(x)\overline{\psi_{p}(y)}\varphi(y)\,.\]
The probability distribution for its possible values over a state
$\psi$ is
\[\Pi_{\psi}(p)=\left|\int_{0}^\infty{\rm
d}x\,\overline{\psi_{p}(x)}\psi(x)\right|^2\,.\]
Notice that this indeed satisfies all the requirements for it to be a
probability.

Let us see in more detail the relationship between the probability
density and the POVM. We have written above the expression for the
action of $F([a,b])$. Symbolically, we can write $F({\rm d}p)$, and
its expectation value on a state $\psi$ is
\begin{eqnarray}
\langle\psi|F({\rm
d}p)\psi\rangle & = & \int_{0}^\infty{\rm
d}x\int_{0}^\infty{\rm
d}y\,\overline{\psi(x)}\,\frac{e^{ip(x-y)/\hbar}}{2\pi\hbar}\psi(y)\,{\rm
d}p\nonumber\\
& = & \Pi_{\psi}(p){\rm d}p\,.\nonumber
\end{eqnarray}
The expression just written defines in fact
$\langle\phi|F({\rm d}p)\psi\rangle$, and therefore the whole POVM.
Furthermore, the domain of the operator associated with the POVM is
also defined by $\Pi_{\psi}(p)$: by integration by parts it can readily be
seen
that the domain of the operator $\hat p$ is precisely the set of
states $\psi$ for which the second moment of the probability
distribution $\Pi_{\psi}(p)$ is finite, and there it coincides with $\|\hat
p\psi\|^2$.
Therefore, the probability density written above defines the POVM and
the operator with it associated. Notice
also
that the
probability density need not be a continuous function. The only
requirement is that $\Pi_{\psi}(p){\rm d}p$ be a good measure on the
real line.

Another important property of the POVM is covariance under
displacements of momenta, which reflects the commutation relation $[\hat
x,\hat
p]=i\hbar$. Namely, it is readily computed that, for all real $q$,
\[\langle \phi|e^{iq\hat x/\hbar}F([a,b])e^{-iq\hat x/\hbar}\psi\rangle =
\langle \phi|F([a+q,b+q])\psi\rangle\,.\]
In terms of the probability density, the statement is that
\[\Pi_{\psi_{q}}(p)=\Pi_{\psi}(p+q)\,,\]
where $\psi_{q}=e^{-iq\hat x/\hbar}\psi$ is the shifted state.

The probability density, written above for the
case of pure states, can be easily generalized to mixed states. Let
$\rho(y,x)$ be the matrix elements of the density matrix $\hat\rho$ in
the position representation. The probability density associated with
this density matrix is then
\[\Pi_{\rho}(p)=\int_{0}^\infty{\rm d}x\int_{0}^\infty{\rm d}y\,
\frac{e^{ip(x-y)/\hbar}}{2\pi\hbar} \rho(y,x)\,.\]
In what follows we will use only pure states in the discussion,
bearing in mind that the generalization to mixed states is
straightforward.

It would seem that something akin to the spectral theorem of
self-adjoint operators has been achieved, and this is indeed the case.
The difference, however, lies in that the expectation value of higher
order powers of $\hat p$ on a state $\psi$, $\langle\psi,\hat
p^n\psi\rangle$, does not necessarily coincide with the corresponding
moment of the distribution $\Pi_{\psi}(p)$, i.e.,
$\int_{-\infty}^\infty{\rm d}p\,p^n \Pi_{\psi}(p)$, for $n\geq3$.  For
the case at hand, for instance, this happens already for $n=3$ if
$\psi'(0)\neq0$.  As a matter of fact, in such a situation the
expectation value for $\hat p^3$ has an imaginary part.  Nonetheless,
the measures of momenta can be readily associated with
$\Pi_{\psi}(p)$, which carries the relevant physical information.

Even so, we would normally like to understand the constructions above
in terms of the more usual recipes for self-adjoint operators.  We
shall now make essential use of the uniqueness theorem for POVMs of
maximally symmetric operators: given a maximally symmetric operator
$\hat A$ over a Hilbert space ${\cal H}$, there is a unique POVM,
$F_{A}$ (unique up to isomorphisms), such that its first operator
moment coincides with the operator, and that the set of states over
which the second moment exists is precisely the domain of $\hat A$
\cite{bglbook,akgl}.  This means that if we are to construct by
whichever means a POVM such that it fulfills these conditions, we will
be obtaining again the same POVM.

Moreover, Naimark's dilation theorem tells us that any POVM
associated with a symmetric operator defined on a dense subset of
$\cal H$ can be constructed from a
self-adjoint
extension of the operator to a larger space $\tilde{\cal H}$, as
follows: let $E$ be the projection valued measure of the self-adjoint
extension (i.e., a POVM that satisfies the further requirement that
$E([a,b])^2=E([a,b])$), and $P$ the projection operator from
$\tilde{\cal H}$ to $\cal H$. Then $F([a,b]):=P E([a,b])$ is a POVM
associated with the symmetric operator.

As a consequence, if we are to build a self-adjoint extension of $\hat
p$ in a larger space, we reproduce the unique POVM and the whole of
the physical content of the operator from the usual analysis for the
self-adjoint extension and a projection. Notice however that the
possible extensions are infinite. Not so the POVM, and that makes the
freedom of choice of extension even more interesting.

Back to the case of $\hat p=-i\hbar\partial_{x}$ defined on the
half-line, we see that there is a
simple possibility: to extend the
operator to the full line, i.e., $-i\hbar\partial_{x}$ defined on a
dense subset of ${\cal H}=L^2({\bf R},{\rm d}x)$. Naturally enough,
the action
of this operator on the elements of $D(\hat p)$ is the same as that
of $\hat p$, so it is an extension.
This is not the only self-adjoint extension on the whole real line,
of course. As a matter of fact, the deficiency indices for the direct
sum of the momentum operators on the positive and negative half-lines
are $(1,1)$, thus signalling that a one dimensional continuum of
alternative self-adjoint extensions exist. They differ by the
presence of a jump function located on $x=0$. However, most natural is
the momentum on the full line, defined over absolutely continuous
functions, i.e. with no jump on $x=0$.
This is a self-adjoint operator, for which the standard spectral
analysis is applicable.

More concretely, the projection valued measure $E$ is given by the
expression
\[\langle \phi|E([a,b])\psi\rangle = \int_{a}^b{\rm
d}p\int_{-\infty}^{\infty}{\rm
d}x \int_{-\infty}^{\infty}{\rm
d}y\,\overline{\psi_{p}(y)}\psi_{p}(x)\overline{\phi(x)}
\psi(y)\,,\]
with $\psi_{p}(x)=\exp(ipx/\hbar)/\sqrt{2\pi\hbar}$, as before, but
now defined over the whole real line.
The projection $P$ in our case
is simply $(P\psi)(x)=\Theta(x)\psi(x)$, with $\Theta$ being
Heaviside's step function.

The probability distribution for the momentum operator over the whole
real line for a state $\psi$ is of course the modulus squared of the
wave function in the momentum representation, and its restriction to
states that belong to $D(\hat p)$ is none other than the probability
distribution associated with the POVM.

We see then that the reason we could not do the standard analysis for
the momentum operator on the half-line is that we are being, in a way,
far too restrictive in the behaviour near $x=0$ of the states on which
it can act.  There is a reminder of the full line, seen for instance
in the principal part that forbids orthogonality, or in the fact that
the higher moments of the probability distribution do not in general
agree with the expectation value of powers of the restricted momentum
operator.  If our states were such that the function $\psi$ and all of
its derivatives were zero at $x=0$, there would be no problem with the
higher moments, and we would get no imaginary part for the powers of
$\hat p$ over such states.
However, such a strong restriction on the
allowable wavefunctions would cut out many physically sensible cases.

Let us now write slightly more formally the extension procedure we
have performed to reobtain and interpret the probability density: we
have started with a maximally symmetric operator $\hat
p=-i\hbar\partial_{x}$ acting on
the dense domain $D(\hat p)\subset{\cal H}_{>}=L^{2}\left({\bf
R}^{+},{\rm d}x\right)$, with deficiency indices $(1,0)$.  We have
then considered an extension in $L^{2}\left({\bf
R},{\rm d}x\right)$, making use of the isomorphism
\[L^{2}\left({\bf
R},{\rm d}x\right)=L^{2}\left({\bf
R}^{+},{\rm d}x\right)\oplus L^{2}\left({\bf
R}^{-},{\rm d}x\right)\,.\]
The extension has been the natural one, i.e., $-i\hbar\partial_{x}$ on
the full line (although, as we repeatedly stated, this is not the only
possible extension; another easy choice would have been some extension
of $\hat p\oplus(-\hat p)$ , for instance).
The standard spectral analysis for this operator
produces the probability density, which, when restricted to ${\cal
H}_{>}$, gives $\Pi_{\psi}(p)$, the probability density out of which
the POVM for the operator we started from can be built.

\section{The time-of-arrival operator}
Within the long-running discussion on the concept of time in quantum
mechanics, and in particular with respect to the status of the
time-energy uncertainty relations, Aharonov and Bohm introduced in an
important paper \cite{ahb}, among other questions,
``a clock'' to  measure time from the position and momentum of a
freely moving
test particle. The corresponding operator was obtained by a simple
symmetrization of the classical expression $m y/p_y$,
where $y$ and $p_y$ are, respectively,
the position and momentum of the test particle. By the same token,
the operator obtained by symmetrizing
the classical expression for the arrival time at $x=0$
of a freely moving particle having position $x$ and momentum $p$, $t=-mx/p=
$,
is given by \cite{microst,mlppra}
\[\hat T_{AB}:= -\frac{m}2\left(\hat x\hat p^{-1} + \hat p^{-1}\hat
x\right)\,,\]
Note the minus sign in comparison to the clock time in \cite{ahb}.
In spite of the somewhat subtle difference (in concept and sign)
with the original
time operator introduced in \cite{ahb} we shall refer to this operator
as the Aharonov-Bohm (time-of-arrival) operator.
Without regard for topological considerations, it is clear that it
has the correct commutation relation with the free particle
Hamiltonian on the line, $\hat H_{0}=\hat p^2/2m$. It is thus a good
candidate for a time operator, with the plausible physical
interpretation, given by the correspondence rule, that it is related
to the time of arrival (notice that other quantization rules might
possibly give a different result, although this is the operator
obtained not just by the symmetrization rule, but also by applying
Weyl, Rivier, or Born-Jordan quantizations \cite{samupa}). However,
it also follows, from Pauli's theorem, that it cannot be a
self-adjoint operator acting on a dense space.

Let us then examine first the question of the domain of $\hat T_{AB}$.  In
order to do that, it is useful at this point to consider the Hilbert
space of the free particle in the momentum representation, ${\cal
H}_{p}:=L^{2}\left({\bf R},{\rm d}p\right)$ (on this space we know how
$\hat p^{-1}$ acts, thanks to the spectral theorem).  Formally, we
then obtain
\[\hat T_{AB}\to\frac{i\hbar m}2\left(\frac1{p^2}-\frac2p
\frac{\partial}{\partial p}\right)\,.\]
There are other alternative expressions, such as $-i\hbar m
p^{-1/2}\partial_{p}p^{-1/2}$, which would be valid for $p>0$, or for
$p<0$ by analytic continuation,
see e.g. \cite{grt}.
At any rate, $\hat T_{AB}$ understood as a
differential operator presents a singular point at $p=0$, and this is
the source of all the difficulties that have appeared in the
literature.

The differential operator written above can only be applied to
absolutely continuous functions, but there are further requirements.
One such is that $\hat T_{AB}\psi$ belongs to ${\cal H}_{p}$, i.e., that it be
square integrable.  This poses a restriction due to the singularity of
the operator at $p=0$.  On computation, one finds that the singularity
is avoided if the function $\psi$ has one of the following possible
behaviours close to $p\to0$: either $\psi(p)\sim p^{1/2}$, or
$\psi(p)/p^{3/2}\to0$.  However, this is not enough to fix the domain
of the operator, as Paul noticed long ago \cite{paul}.  Given that, at
least formally, $\hat T_{AB}$ is symmetric, this should also be a requirement
on its domain.  Integration by parts, and demanding that
$\langle\varphi|\hat T_{AB}\varphi\rangle$ be equal to
$\langle\hat T_{AB}\varphi|\varphi\rangle$ for all $\varphi$ in the domain of
$\hat T_{AB}$, leads us to exclude the first possibility, thus defining the
domain of $\hat T_{AB}$, $D(\hat T_{AB})$, as the set of absolutely continuous square
integrable functions of $p$ on the real line, such that
$\psi(p)/p^{3/2}\to0$ as $p\to0$ and $\|\hat T_{AB}\psi\|^2$ is finite.
As for the alternative expression
$-i\hbar mp^{-1/2}\partial_{p}p^{-1/2}$, let us ask that $(\Theta(p)-
i\Theta(-p))\psi(p)/\sqrt{|p|}$ be absolutely continuous for
$\psi$ to be in its domain. This, together with the further
requirement of symmetry, leads us to a domain that
coincides with that of $\hat T_{AB}$.
Since the respective actions also
coincide over this domain, we see that they are but equivalent
differential expressions for the operator, once the adequate domain is
taken into account.

$\hat T_{AB}$
is closed over this domain, as can be checked by computing
$\left(\hat T_{AB}^\dagger\right)^\dagger$ (there is no doubt that $D(\hat T_{AB})$ is
dense in ${\cal H}_{p}$, so the closure of $\hat T_{AB}$ must coincide with
$\left(\hat T_{AB}^\dagger\right)^\dagger$).  Before that, though, let us
examine $\hat T_{AB}^\dagger$ and its domain.

In order to apply von Neumann's formula, we have to check whether
there are states $\psi$ in ${\cal H}_{p}$ such that for all
$\varphi\in D(\hat T_{AB})$ the following expression holds:
\[\langle\psi|\left(\hat T_{AB}+i\right)\varphi\rangle=0\,,\]
since then $\psi$ is an eigenvector of $\hat T_{AB}^\dagger$ with eigenvalue
$i$.  Analogously, we also have to study the case with eigenvalue
$-i$.  By integration by parts, and application of the condition that
all functions in $D(\hat T_{AB})$ satisfy, namely, that $\varphi(p)/p^{3/2}$
tends to zero as $p$ tends to zero, it is found that there are {\sl
two} independent eigenvectors with eigenvalue $i$, and none with
eigenvalue $-i$.  The relevant eigenvectors are
\[\psi_{\pm}(p)=\Theta(\pm p)\, \sqrt{\pm p}\,
e^{-p^2/2m\hbar}\,.\]
Notice that, contrary to the expectations of some authors
\cite{razavi,grt,oru}, both of them have to be taken into account:
there are no requirements of derivability or continuity for the
functions in the domain of the adjoint.

The deficiency indices are therefore $(2,0)$, and we have a maximally
symmetric operator.  As we have seen in the previous section, this
implies that no self-adjoint extension can exist, and we should redo
for this case the analysis performed for the momentum operator on the
half-line. However, the most convenient way of doing that is by
passing to the energy representation, as was already pointed out by
Allcock in his seminal work \cite{allcock}, and emphasized again by
Kijowski \cite{kijold,kijnew}.
This change of representation is
useful, from the mathematical point of view, because it implements a
theorem \cite{akgl} which states that a simple symmetric operator with
deficiency indices $(n,0)$ can be decomposed as a direct sum of $n$
operators with deficiency indices $(1,0)$. Since each of these is in fact
isomorphic to the momentum operator on the half-line, we will be able
to use directly the previous results.

The change of representation corresponds to the decomposition of the
Hilbert space ${\cal H}_{p}$ into the subspaces of positive and
negative momentum, i.e.,
\begin{eqnarray}
 L^{2}\left({\bf R},{\rm d}p\right)
 & = & L^{2}\left({\bf R}^{+},{\rm
d}E\right)\oplus L^{2}\left({\bf R}^{+},{\rm
d}E\right)\nonumber
\\
& = & {\cal H}_{+}\oplus{\cal H}_{-}\,,\nonumber
\end{eqnarray}
where the first subspace is that of positive momenta, whereas the
second corresponds to negative $p$.  The explicit isomorphism is given
by
\begin{eqnarray}
\psi_{\pm}(E) & = & (m/2E)^{1/4} \psi(\pm\sqrt{2m E})\,,\nonumber\\
\psi(p) & = &
\left(\frac{|p|}{m}\right)^{1/2}
\left[\Theta(p)\psi_{+}\left(\frac{p^2}{2m}\right) +
\Theta(-p)\psi_{-}\left(\frac{p^2}{2m}\right)\right]\,,\nonumber
\end{eqnarray}
where $\psi\in{\cal H}_{p}$,
$\psi_{\pm}\in{\cal H}_{\pm}$, and the isomorphism relates
$\psi\leftrightarrow(\psi_{+},\psi_{-})$.  The factor $E^{-1/4}$ is
due to the change in the measure from ${\rm d}p$ to ${\rm d}E$ and
reciprocally for the factor $\left(|p|/m\right)^{1/2}$.  The
interesting point is that given this isomorphism, the time operator of
Aharonov and Bohm takes the form $-i\hbar\partial_{E}$, as is well
known.  The domain of the operator, $D(\hat T_{AB})$ is sent by the isomorphism
into the direct sums of square integrable absolutely continuous
functions in each subspace, such that for each subspace we have the
restriction that $\psi_{\alpha}(E) E^{-1/2}\to0$ as $E\to0$.  But this
is exactly the case considered above, since the requirement
$\psi_{\alpha}(0)=0$ and the square integrability of
$\psi_{\alpha}'(E)$ imply the restriction stated before.  In other
words, we have the isomorphism
\[\hat T_{AB}=(-i\hbar\partial_{E})\oplus (-i\hbar\partial_{E})=
\hat T_{+}\oplus\hat T_{-}\,,\]
where  $\hat T_{\pm}$ are isomorphic to the momentum operator on the
half-line.

Therefore, the constructions carried out in the previous
section can be
immediately translated to this situation, but taking into account that
the energy spectrum is degenerate whereas the position spectrum is
not. For instance, the complete non-orthogonal set of generalized
eigenfunctions $\psi_{p}$ is doubled here into a set with a continuous
parameter $t$ (the notation is intended to be suggestive) and a
discrete one, with values $+$ or $-$, as follows:
\[\psi_{+}^{(t)}(E)
=\left(\frac{1}{\sqrt{2\pi\hbar}}\,e^{iEt/\hbar},0\right)\]
and
\[\psi_{-}^{(t)}(E)=
\left(0,\frac{1}{\sqrt{2\pi\hbar}}\,e^{iEt/\hbar}\right)\,.\]
In what follows we will not make the explicit distinction between the
element of the full Hilbert space and its component, if only one is
zero.

These functions transform under the isomorphism to give the following
expressions:
\[\tilde\psi^{(t)}_{\alpha}(p)=\Theta(\alpha p)\left(\frac{\alpha
p}{2\pi m \hbar}\right)^{1/2} e^{i p^2 t/2 m\hbar}\,.\]
It is straightforward to prove completeness, i.e.,
$\sum_{\alpha}\int_{-\infty}^\infty{\rm
d}t\,\tilde\psi_{\alpha}^{(t)}(p')\tilde\psi_{\alpha}^{(t)}(p)
=\delta(p-p')$.
Alternatively,
$\sum_{\alpha}\int_{-\infty}^\infty{\rm
d}t\,\psi_{\alpha}^{(t)}(E')\psi_{\alpha}^{(t)}(E)=\delta(E-E')\openone$,
where $\openone$ is the two by two identity matrix
($\delta(E-E')\openone$ is the identity operator on the full Hilbert
space ${\cal H}_{+}\oplus{\cal H}_{-}$).

Nonorthogonality is also a direct translation:
\[\int_{0}^\infty{\rm
d}E\,\psi^{(t')}_{\alpha'}(E)\psi^{(t)}_{\alpha}(E)=\frac12
\delta_{\alpha\alpha'}\left(\delta(t-t')+\frac{i}{\pi}{\rm
P}\frac{1}{t-t'}\right)\]
or
\[\int_{-\infty}^\infty{\rm
d}p\,\tilde\psi^{(t')}_{\alpha'}(p)\tilde\psi^{(t)}_{\alpha}(p)=\frac12
\delta_{\alpha\alpha'}\left(\delta(t-t')+\frac{i}{\pi}{\rm
P}\frac{1}{t-t'}\right)\,.\]

It now behooves us to compute the POVM or, alternatively, the
probability distribution for measured values of the $\hat T_{AB}$ operator
from which it can be readily recovered. By
direct translation, we can write the probability density as
\begin{eqnarray}
	 \Pi_{(\psi_{+},\psi_{-})}(t)& = & \left|\int_{0}^\infty{\rm
d}E\,\frac{e^{-iE
t/\hbar}}{\sqrt{2\pi\hbar}}\psi_{+}(E)\right|^2 +\nonumber
	\\
	 &  & \quad
\left|\int_{0}^\infty{\rm
d}E\,\frac{e^{-iE t/\hbar}}{\sqrt{2\pi\hbar}}\psi_{-}(E)\right|^2
\,,\nonumber
\end{eqnarray}
in the energy representation, or as
\begin{eqnarray}
\Pi_{\psi}(t) & = & \left|\int_{0}^\infty{\rm d}p\,\left(\frac{p}{2\pi
m\hbar}\right)^{1/2} e^{-ip^2 t/2m\hbar}\psi(p)\right|^2
+\nonumber\\ & &\quad
\left|\int_{-\infty}^0{\rm d}p\,\left(\frac{-p}{2\pi
m\hbar}\right)^{1/2} e^{-ip^2 t/2m\hbar}\psi(p)\right|^2\,,\nonumber
\end{eqnarray}
in the momentum representation. This is, of course, the same as
Kijowski's probability density. The essential property of covariance
under transformations generated by the Hamiltonian is also evident and
a direct translation of the covariance property signalled in the
previous section.  Physically, it means that the probability of
arriving at $t$ for a given state is equal to the probability of
arriving at $t-\tau$ for the same state evolved a time $\tau$.
This is the reflection on the probability density of the canonical
commutation relation $[\hat H,\hat T_{AB}]=i\hbar$.

The domain of $\hat T_{AB}$, now defined through the POVM, can be characterized
as the set of elements of the Hilbert space for which
$\int_{-\infty}^\infty{\rm d}t\,\Pi_{\psi}(t) t^2$ is finite, and
this quantity defines $\langle\hat T_{AB}\psi|\hat T_{AB}\psi\rangle$ (thus realizing
the minimum variance demanded by Kijowski
and Werner \cite{kijold,werner}).

As before, we would like to understand all these constructions in
terms of a generalized self-adjoint extension, by using the uniqueness
of the POVM associated with a maximally symmetric operator, and
Naimark's theorem.  {}From the structure of the time operator of
Aharonov and Bohm, namely $\hat T_{AB}=(-i\hbar\partial_{E})\oplus
(-i\hbar\partial_{E})$ on $L^{2}\left({\bf R}^{+},{\rm
d}E\right)\oplus L^{2}\left({\bf R}^{+},{\rm d}E\right)$, it follows
that a {\sl natural} and simple
extension (natural in the sense of following
the natural extension in the analogy) is the operator
$(-i\hbar\partial_{E})\oplus (-i\hbar\partial_{E})$ acting on
$L^{2}\left({\bf R},{\rm d}E\right)\oplus L^{2}\left({\bf R},{\rm
d}E\right)$, which is obviously self-adjoint.  In other words, we have
introduced negative energies, respecting the twofold degeneracy of the
initial spectrum.  On this space the (doubled) Fourier transform acts
as a unitary transformation that provides us with the {\sl time}
representation of the states, and the probability density for the time
of arrival over a pure state in this extended space is nothing but the
modulus squared of the wavefunction in the time representation.
Notice that the time representation corresponds to the doubled space
$L^{2}\left({\bf R},{\rm d}t\right)\oplus L^{2}\left({\bf R},{\rm
d}t\right)$.  We reobtain
the by now usual probability density for
time of arrival, by restriction to the initial Hilbert space.  The
restrictions to the initial Hilbert space of the applications of the
spectral theorem are therefore related to $\Pi_{\psi}(t)$, as stated
before.

The problems that the vicinity of $p=0$ (alternatively $E=0$) pose for
the analysis of the operator of Aharanov and Bohm are thus seen to be
due to the (physically imposed) restriction to the space of positive
energies. There is a reminder of the 
negative energies in aspects such as the nonorthogonality of the set of
generalized eigenfunctions, similarly to what
happens in the paradigmatic example of the previous section.

\section{General time operators}

In previous sections we have discussed the time of arrival for free motion
but in fact its mathematical form, and its
POVM, can easily be generalized for arbitrary potential functions.
The important point however is that in general the resulting
operators and POVMs cannot be physically associated with an
arrival time.

Consider a self-adjoint Hamiltonian, bounded from below, defined on a
Hilbert space $\cal H$, such that the spectral decomposition leads
to the isomorphism
\[{\cal
H}=\oplus_{i=1}^{N}\left(\oplus_{j=1}^{n_{i}}
L^2\left([a_{i},\infty),{\rm d}\mu_{i}(E)\right)\right)\,,\]
where all the measures $\mu_{i}$ are absolutely continuous with
respect to the Lebesgue measure. That is to say, the spectrum of the
Hamiltonian is absolutely continuous, and for any given $E\in{\bf R}$
the degeneracy is finite, the maximum degeneracy being
$\sum_{i=1}^{N}n_{i}$. The Hamiltonian is realized in each subspace
as the multiplication by the variable $E$. Therefore, a natural (but
most definitely not unique) candidate for the time operator is
again $-i\hbar\partial_{E}$ on each of these subspaces, restricted to
functions that vanish at $a_{i}$. Again each
component of this time operator can
be extended to the real line, where it will be self-adjoint and admit
a spectral decomposition, which, by projection, will give us again
the POVM associated with our candidate operator. The POVM, and its
associated probability density, will be unique for the given operator
if this is indeed maximally symmetric.

Notice moreover that if we do extend one of the Hilbert subspaces to
be square $\mu$-integrable functions on the real line, and demand that
the covariance property of the restriction of the time operator to
that Hilbert subspace be mantained on the real line, then, under
fairly general conditions, the extension of  time operator and
Hamiltonian on $L^2\left({\bf R},{\rm d}\mu_{i}(E)\right)$ will be
unitarily equivalent to the canonically conjugate pair
$-i\hbar\partial_{E}$ and multiplication by $E$. Therefore, modulo
intervening unitary transformations, we have a way of constructing
POVMs for time operators by Naimark dilations.

The usefulness or otherwise of the particular time operator under
consideration will then have more to do with its actual properties and
measured probability density rather than with mathematical
difficulties that had riddled the work of many previous workers in the
field.

To illustrate this point further, let us examine the standard example
of a particle in a constant field. The Hamiltonian is $\hat H_{g}=\hat
p^2/2m + mg \hat q$, clearly seen to be unbounded. Pauli's argument
is therefore not applicable, and a self-adjoint operator canonically
conjugate to $\hat H_{g}$ does indeed exist, namely, $\hat T_{g}=\hat
p/mg$. In the energy representation this operator is of course given
by $-i\hbar\partial_{E}$. It can be ascribed to time of arrival to
zero momentum, but in no possible way to time of arrival to a
specified position. This example tells us that an operator having the
form $-i\hbar\partial_{E}$ in the energy representation, be it as a
self-adjoint operator, or associated with a POVM, does not necessarily
mean that we have obtained a time-of-arrival operator, even though a
time operator is indeed being considered.

At this point it is worthwhile mentioning the approach of Le\'on et
al. \cite{ljpu}, intended to generalize $\hat T_{AB}$ to the interacting case
of a potential barrier
(under the hypothesis of asymptotic completeness and absence of bound
states). They do obtain good time operators,
involving $-i\hbar\partial_E$ in the basis of outgoing or
incoming scattering states, and their distributions,
understood as POVMs. However, since the expectation value of these
operators is, over a given state, the same as the expected time of
arrival that the corresponding asymptotic states
(outgoing or incoming respectively) would produce in the
case of no interaction,
it is hard to understand them as  bona fide
time-of-arrival operators at an arbitrary spatial point.

Finally, it is also interesting to examine the relation of the operator
$\hat T_{AB}$ with older literature on the ``time operator'' by Olkhovski, Recami,
and others \cite{OR}. The basic idea behind this set of works is
to extract a time operator from the relation
that defines an average ``presence time'' at $x=0$ by
\[\langle t \rangle\equiv \frac{\int_{-\infty}^{+\infty} dt\,
|\psi(x=0,t)|^2
t}{\int _{-\infty}^{+\infty} dt\,|\psi(x=0,t)|^2}\,.\]
We shall discuss as in \cite{OR} the simple case of states without negative
momenta so that there is no need to consider the energy degeneracy
in the following expressions.
The energy Fourier transform of $\psi(x=0,t)$ is given by
\begin{eqnarray}
\eta(E) & = & h^{-{1/2}}\int _{-\infty}^{+\infty} \psi(x=0,t)
e^{iEt/\hbar}\,dt\nonumber
\\
& = & h^{1/2}\langle x=0|E\rangle \langle
E|\psi(t=0)\rangle\,.\nonumber
\end{eqnarray}
In terms of $\eta$ we can
write
\[\langle t \rangle=-i\hbar
\int_0^\infty dE\, \tilde{\eta}(E)^* \frac{\partial \tilde{\eta}(E)}
{\partial E}\,,\]
where $\tilde{\eta}(E)\equiv \eta(E)/(\int dt |\psi(0,t)|^2)^{1/2})$.
Thus a ``time operator'' may be again identified with
$-i\hbar\partial_E$, but now in the space of functions
$\tilde{\eta}(E)$. This operator is different
from $\hat T_{AB}$.

In terms of the position and momentum operators
the average presence time becomes
\begin{equation}
\langle t \rangle
=-\frac{m}2\frac{\langle \psi(t=0)| \hat{p}^{-2}\hat{x}+\hat{x}
\hat{p}^{-2}
|\psi(t=0)\rangle}{\langle \psi(t=0)|
\hat{p}^{-1}|\psi(t=0)\rangle}\,,\label{presencia}
\end{equation}
provided the integrals exist.
This is to be contrasted with the average time that can be defined using
the current density $J(x=0,t)$,
\begin{equation}
\langle t \rangle_J\equiv \frac{\int_{-\infty}^{\infty} dt\,J(x=0,t) t}
{\int_{-\infty}^\infty dt\,
J(x=0,t)}=\langle \psi(t=0)|\hat T_{AB}|\psi(t=0)\rangle\,.\label{llegada}
\end{equation}
The different formal results are to be expected since,
in the classical limit,
(\ref{presencia}) is associated with an average presence time and
(\ref{llegada}) with an
average passage time.  Note however that in both cases we can
relate these measurements with the formal operator
$-i\hbar\partial_{E}$, as has been pointed out all along.

As stated above, no matter how one constructs general time operators,
one is inevitably led to $-i\hbar\partial_{E}$ (plus, possibly,
a function of $E$), but the physical interpretation of those
operators, self-adjoint or associated with a POVM, is quite another
issue.

\section{Discussion}
In the sections above we have made reference to results and proposals
of other workers in the subject. In this section, however, we intend
to restate some of those results and discussions in the light of our
description. To start with, none better than the initial paper of
Aharonov and Bohm \cite{ahb}, where they introduce the operator $\hat T_{AB}$.
They were not particularly concerned with the self-adjointness or
otherwise of the operator, but they did signal the problem of the
singularity for $p=0$, which they brushed aside by stating that all
components of the interesting wavefunctions would be of high enough
momentum. Insofar as that is the case, then there is indeed no
problem in obtaining a probability distribution for measurements of
the operator.
Their real point of interest in
introducing $\hat T_{AB}$ was to show that there is no reason inherent in the
principles of quantum mechanics for the energy of an observed system
not to be observed in as short a time as one pleases, where the time
is measured in an observer system.
The topic of measurement limitations is retaken in much later
work of Aharonov and
others \cite{aopru}, where they make the distinction between direct
and indirect measurements of time of arrival, indirect measurements
being associated with $\hat T_{AB}$ (or with regularized variants),
and direct measurements with couplings of the particle with other
degrees of freedom that in the classical limit would measure
the time of arrival.
They describe a lower bound for the time uncertainty of direct measurements
inversely proportional to the typical energy of the particle
(not to the energy uncertainty).
In fact the same dependence has been recently described, and
traced back to the lower bound of the energy
spectrum, for the indirect measurement case \cite{lett}. But
the time uncertainty refers to an ensemble and does not
preclude a precise determination of the time of arrival
in an individual measurement, in other words, this does not
involve a limitation of the quantum theory to handle the
arrival time concept and prescribe {\it intrinsic} arrival time
distributions (formulated without explicit recourse to a
measuring apparatus or an additional ``clock'' degree of freedom).
Numerical results have shown
essential agreement between the (intrinsic) POVM
distribution and  operational distributions given by
phenomenological screen models,
except in rather pathological cases \cite{mbm,mpl}.

Going back in time again, we come to the work of Allcock
\cite{allcock}.  As we have signalled above, he advocated the use of
the energy representation. Even more, he signalled the need for
negative energies in a particular way. If we indeed want to measure
arrival times, we would have to prepare wavepackets that are localized
on one side of the point of arrival ($x=x_{0}$, say) for all times before
a given one, $t<t_{0}$. This directly entails that the temporal
Fourier transform of the wavefunction will exhibit all positive and
negative energies, save for a set of null measure. In a way, it could
be argued that the conceptual setup for measuring time of arrival
requires by itself the introduction of negative energies.
Furthermore, the proof he proposes for the need for all positive and
negative energies is also illuminating: it goes the way of the
Paley-Wiener-Titchmarsh theorem and the uniqueness theorem for
analytic continuations. This is precisely the same route taken to show
the non-orthogonality of any set of functions that could be considered
as (generalized) eigenfunctions of a time operator. In that proof it
is clear that the introduction of negative energies also eliminates
the problem of non-orthogonality.  However, his interpretation of the
non-orthogonality of (generalized) eigenfunctions for the time of
arrival as precluding any measurement thereof is at odds with the
operational approach within which the POVM idea takes place.

A number of more mathematical papers has also been published, with a
more functional analytical approach than those mentioned before
\cite{paul,razavi,gianni}.

The axiomatic approach of Kijowski \cite{kijold} has been mentioned
all along, since its guiding idea, i.e., characterizing the physical
content of a concept such as time of arrival in a probability
distribution, is crucial for our purposes, and clears the path for the
introduction of the POVM as the physically relevant object.  Kijowski
also emphasized the usefulness of the energy representation, and
decomposed the Hilbert space of the free particle as the direct sum of
the subspaces of positive and negative momenta.  He constructs both
the operators $\hat T_{\pm}$ (in his notation, $\hat t^{Q}_{\pm}$, if
$Q$ is the point $x_{3}=0$) by comparison with the probability
distribution, as we do. He then points out that $\hat T_{+}\oplus\hat
T_{-}$ admits no self-adjoint extension, and goes no further that way.
However, he also introduces a particularly interesting object, retaken
more recently by Delgado and Muga \cite{delmu} (see also
\cite{kijnew}), namely $\hat T'=\hat T_{+}\oplus\left(-\hat
T_{-}\right)$. He then asserts that this operator is essentially
self-adjoint.  Let us examine this question from the point of view
taken in this paper.  As we have signalled, both $\hat T_{+}$ and
$\hat T_{-}$ are isomorphic to the momentum operator on the half-line,
and their deficiency indices are $(1,0)$ for both of them.  Therefore,
$-\hat T_{-}$ is also a maximally symmetric operator, with deficiency
indices $(0,1)$.
The total operator $\hat T_{+}\oplus\left(-\hat
T_{-}\right)$, symmetric on a
dense subspace, has deficiency indices $(1,1)$, so it is not
self-adjoint, although it admits self-adjoint extensions.
Contrary to what might be inferred from some claims in \cite{kijnew},
there are infinite
self-adjoint extensions, parametrized by $\alpha\in[0,2\pi)$. The
extension $\hat T'_{\alpha}$ is defined on
$D(\hat T'_{\alpha})$, which is the set of elements of ${\cal
H}_{+}\oplus{\cal H}_{-}$ of the form $\left(\psi_{+}+\lambda
e^{-E/\hbar},\psi_{-}+\lambda e^{i\alpha}
e^{-E/\hbar}\right)$,
where $\psi_{\pm}\in D(\hat T_{\pm})$, and $\lambda\in{\bf
C}$. The action on its domain is simply that of the couple of
differential
operators $(-i\hbar\partial_{E},i\hbar\partial_{E})$. The boundary
conditions, however, are different. In terms of the momentum
representation, the functions in the domain of $\hat T'_{\alpha}$ can
have $p^{1/2}$ behaviour, for instance. At any rate, the probability
distribution deduced from these operators is not covariant under time
shifts, so their physical interpretation is rather clouded
unless the state belongs to only one of the two subspaces (positive
or negative momenta)
\cite{mlppra}.

Another interesting attempt to construct a self-adjoint operator
related to the time of arrival is that of Grot, Rovelli and Tate
\cite{grt}. In essence, they give a realization of the intuition of
Aharonov and Bohm \cite{ahb} that for high momenta the $p=0$
singularity of the differential operator is irrelevant, a point later
taken up by Paul \cite{paul} as well.  However, this generalized
self-adjoint operator has also some drawbacks from the point of view of
its interpretation \cite{mlppra,oru}.
{}From the point of view taken in this paper, the problem with the
proposal of Grot, Rovelli and Tate is that the strong operator limit
of their deformed operator does not exist when their regulator is made
to disappear (this can be seen, for example, computing the norm of the
action by the regulated operator on any gaussian state, and then
trying to eliminate the regulator - the norm blows up).  Therefore,
the results obtained with the regulator cannot be at all extrapolated
to the unregulated case.

Closest to the work carried out in this paper is the presentation by
Busch, Grabowski and Lahti \cite{bglbook,bglpaper}. However, they
refer the idea of using Naimark's dilation in this context to the
paper of Werner\cite{werner}, where he constructs ``screen observables''
using a
version of imprimitivity adequate for the case of POVMs. Our analysis
is much more simplified and explicit.

In conclusion, we have given a description of the POVMs associated
with the momentum operator on the half-line and the time operator of
Aharonov and Bohm in terms of Naimark's dilations. More concretely,
in terms of what for the momentum operator is the natural dilation. We
have clarified several issues that have been the matter of debate
or confusion
in the literature as of late, in particular the deficiency indices
and constructions of self-adjoint extensions of $\hat T_{AB}$, and we have
rederived the probability density for times-of-arrival derived
axiomatically by Kijowski. We have also made a proposal for
the construction of the probability distributions of a
wide class of time operators,
for which
it has been shown
that they are not in general related to the arrival time, in spite of the
formal agreement with the arrival time operator of the free motion case.
\acknowledgements
We thank W. Amrein for very useful comments.
Support by the
CICYT under projects AEN-96-1668 and PB87-1482, by Gobierno
de Canarias  (Grant No. PB/95), and by the University of the Basque
Country (Grant No. UPV063.310-EB187/98) is
gratefully acknowledged.

\end{document}